# Cavity nano-optomechanics: a nanomechanical system in a high finesse optical cavity

Sebastian Stapfner[a*], Ivan Favero[b†], David Hunger[a], Philipp Paulitschke[a], Jakob Reichel[c], Khaled Karrai[a], Eva M. Weig[a‡]

[a]Ludwig-Maximilians-Universität München, Center for Nano Science and Fakultät für Physik, Geschwister-Scholl-Platz 1, 80539 München, Germany;
[b]Laboratoire Matériaux et Phénomènes Quantiques, Université Paris-Diderot, CNRS, UMR 7162, 10 rue Alice Domon et Léonie Duquet, 75013 Paris, France;
[c]Laboratoire Kastler Brossel, Ecole Normale Supérieure, Université Pierre et Marie Curie, CNRS, 24 rue Lhomond, 75005 Paris, France

## ABSTRACT

The coupling of mechanical oscillators with light has seen a recent surge of interest, as recent reviews report.[1,2] This coupling is enhanced when confining light in an optical cavity where the mechanical oscillator is integrated as back-mirror or movable wall. At the nano-scale, the optomechanical coupling increases further thanks to a smaller optomechanical interaction volume and reduced mass of the mechanical oscillator. In view of realizing such cavity nano-optomechanics experiments, a scheme was proposed where a sub-wavelength sized nanomechanical oscillator is coupled to a high finesse optical microcavity.[3] Here we present such an experiment involving a single nanomechanical rod precisely positioned into the confined mode of a miniature Fabry-Pérot cavity.[4] We describe the employed stabilized cavity set-up and related finesse measurements. We proceed characterizing the nanorod vibration properties using ultrasonic piezo-actuation methods. Using the optical cavity as a transducer of nanomechanical motion, we monitor optically the piezo-driven nanorod vibration. On top of extending cavity quantum electrodynamics concepts to nanomechanical systems, cavity nano-optomechanics should advance into precision displacement measurements near the standard quantum limit[5], investigation of mechanical systems in their quantum regime, non-linear dynamics[6] and sensing applications.

**Keywords:** cavity optomechanics, nanomechanics, micro-cavity, fiber optics, optical sensing,

## 1. INTRODUCTION

Research in the field of cavity-optomechanics has attracted more and more attention during the last decade. Researchers study the dynamics of deformable Fabry-Pérot cavities in a continuously growing variety of experiments targeting backaction cooling of a mechanical resonator striving for low phonon occupation states towards the quantum regime.[1,2]

---

[*] sebastian.stapfner@physik.uni-muenchen.de
[†] ivan.favero@univ-paris-diderot.fr
[‡] weig@lmu.de

Optomechanical resonators in different set-ups range from gram scale mirrors[7] across micro resonators of the size of AFM-cantilevers[8, 9, 10], micro toroids[11] and spheres[12] to resonators integrated in photonic waveguide circuits[13, 14] with masses down to the picogram scale. Following the derivations of cavity optomechanics[8] the cooling efficiency and the chance to reach the quantum ground-state increase as the mass of the mechanical resonator decreases, as its frequency and quality factor increase and as the finesse of the cavity becomes larger. The first three requirements call for smaller, nanoscale mechanical objects that can no longer be integrated to form a mirror of a high finesse cavity due to increased diffraction losses. Existing concepts could however be extended to this regime.[3] As opposed to recent developments of lithographically integrated optomechanical schemes based on two-dimensional photonic waveguide chips[13, 14], our approach pursues the idea of combining a nanomechanical resonator with a fixed miniature Fabry-Pérot cavity.[3, 4] This approach is extremely versatile, in the sense that basically any kind of nanomechanical resonator can be positioned in the independent cavity. On top of this asset, hybrid systems[4, 15] which separate optical and mechanical components enable discrete tuning of either element while in other experiments[7-13, 16] the integrated optomechanical element needs to exhibit marvellous optical and mechanical properties at the same time.

## 2. OPTICAL CAVITY

A high finesse Fabry-Pérot cavity with extremely small mode volume is realized between two opposing glass fibre end-facets schematically drawn in figure 1a).[17] The input fibre is a single mode fibre (SM) matching to 780 nm whereas the output fibre is a multi mode fibre (MM) with 50 µm core diameter in order to collect as much of the transmitted light as possible. The end-facets are concavely shaped by $CO_2$ laser ablation to form the cavity mirrors. Additional dielectric layers are deposited on them to obtain highly reflective Bragg-mirrors optimized for a wavelength of 780 nm.[17] The two fibre ends are positioned and aligned opposing each other to define a stable optical resonator with a length of 37 µm and an optical mode waist radius of 3.4 µm (see figure 1a and figure 2c). This technology allows for small mode volume optical cavities for which finesses up to $F = 37,000$ have been reported at ambient conditions.[17]

The fibre ends are glued on shear piezo elements (PZT1 and PZT2) in order to allow for a cavity length modulation. By applying voltage ramps between ±410 V to either of the shear piezos the cavity is swept across three Fabry-Pérot resonances covering two free spectral ranges (FSR) of the cavity. The finesse of the cavity can then be measured by keeping the wavelength fixed and by scanning the length of the cavity. Since the cavity length directly depends on the voltage $V_{PZT2}$ applied to PZT2, the finesse is given by $F = v_{FSR}/\Delta v = l_{FSR}/\Delta l = V_{FSR}/\Delta V$ where $v_{FSR}$, $l_{FSR}$ and $V_{FSR}$ denote the shift in optical frequency, cavity length and voltage to cover one FSR of the cavity mode and $\Delta v$, $\Delta l$ and $\Delta V$ are the respective changes in optical frequency, cavity length and ramp voltage $V_{PZT2}$ to scan across a cavity line full width at half maximum as can be seen in the plots in figure 1b) and c).

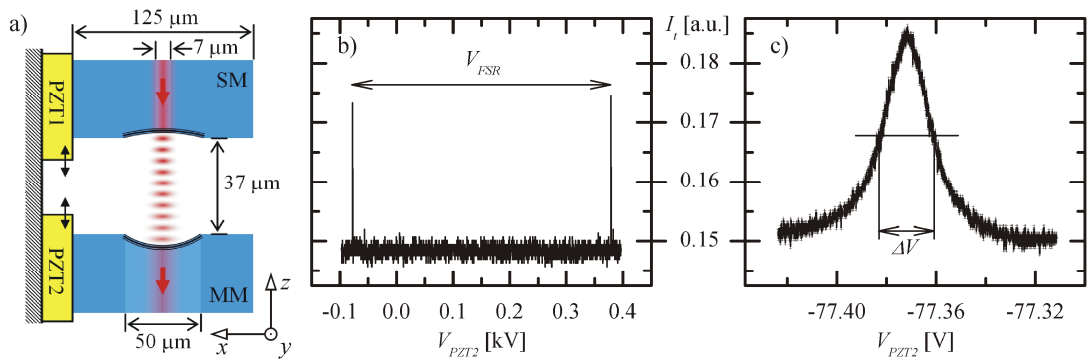

Figure 1. (color online) Cavity and finesse measurement. a) Schematical drawing of the fibre-based cavity. Two opposing fibre ends (SM and MM in blue) both with concave Bragg-layer coated facets form a miniature high finesse Fabry-Pérot cavity. Light (in red) enters and exits the cavity through a single mode fibre (SM) and a multi mode fibre (MM) respectively. The coordinate system helps to orient in figures below. b) Scan of the cavity transmission across one FSR and c) across the left resonance in b). Both graphs show the light intensity transmitted through the cavity in dependence of the voltage applied to the shear piezo (PZT2) moving the rear mirror.

Even though our cavity displays an initial finesse of only 5,000 in air we demonstrate a significant increase of the finesse upon removal of surface adsorbates presumably present on the mirror surfaces. Our cleaning procedure is based on integration of the cavity into a vacuum chamber (shown in figure 2a) which allows operation between ambient pressure and $10^{-5}$ mbar. After initial evacuation to $10^{-5}$ mbar, the cavity finesse is observed to increase to 12,000. After applying several cycles of evacuating and purging the chamber with dry nitrogen at ambient pressure the finesse stabilizes at 21,000.

A second independent method to measure the cavity finesse is implemented to verify the first measurement. To this end the laser is modulated at a frequency $\upsilon_m = 1$ GHz creating sidebands in the laser light spectrum apart from the laser frequency $\upsilon_o$. These sidebands with frequencies $\upsilon_o \pm \upsilon_m$ appear as additional peaks symmetrically located on either side of the cavity resonance peak when scanning the cavity length. The distance between those extra peaks and the cavity resonance peak is known to be $\upsilon_m$, which can be employed to calibrate the scan voltage to frequency units and measuring the cavity line width $\Delta \upsilon$. With the FSR $\upsilon_{FSR} = c/2L$ calculated from the cavity length $L = 37\mu m$, the finesse is 24,500 which is in a good agreement with the first measurement within a measurement imprecision of 10%.

The vacuum chamber consists of a glass cell and standard metal vacuum tubing forming a solid body. The solid body provides feed throughs for electrical wires and glass fibres, access to pressure gauges and the pumping line and forms a solid mount for the cavity and the xyz-positioning unit (attocube, ANP100). Cavity and positioning units are covered by the glass cell (HELLMA, 700.036-OG) to enable visual access (figure 2a). Rapid sample cycling is enabled by connecting the glass cell with only one KF 40 quick flange to the residual vacuum chamber. Figure 2b) and c) show a close-up series of images of the cavity taken with a binocular microscope positioned outside the glass cell.

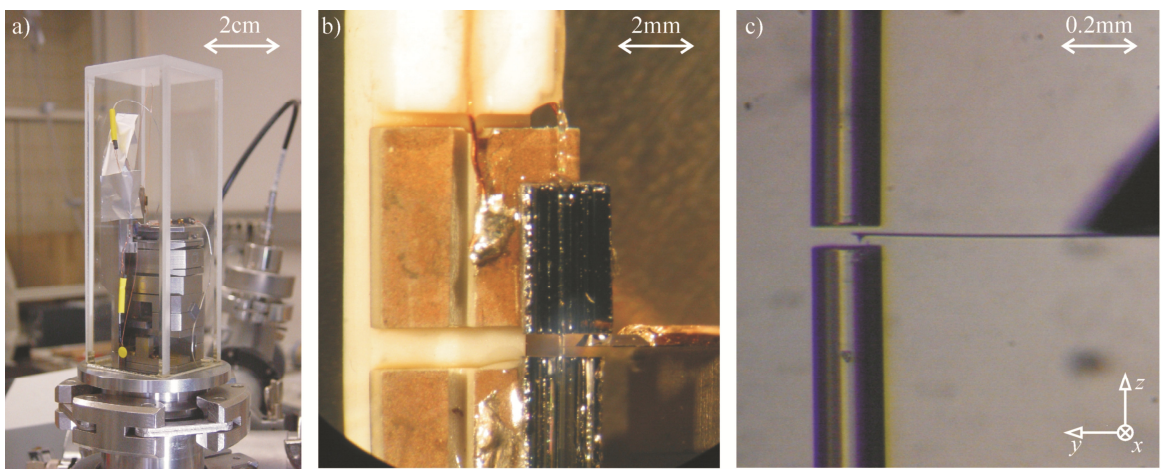

Figure 2. (color online) Cavity set-up close-up series. a) A sealed glass cell allows to operate the cavity in vacuum. It hosts the cavity and the xyz-positioning unit to place the samples precisely in the cavity gap. A microscope (not seen here) provides visual access to the cavity. b) Close-up showing the cavity assembly. Both glass fibers are vertically guided in v-grooves in separate silicon substrates (dark rectangle) that are situated on the shear piezo elements (gold colored). A marcor block (white) forms a solid common basis for both piezo-v-groove-fibre-stacks. The piezos are electrically contacted via soldered copper wires. The silicon chip hosting an AFM-cantilever is placed near the cavity between the v-groove substrates on the right. c) Zoom onto the cavity with the front fibre (SM) coming from above and the rear fiber (MM) from below. An AFM-cantilever hosting a nanomechanical system attached at the lever's end along the lever axis (not visible here, see figure 5a for details) penetrates the gap between the two fibres.

Figure 3 depicts the experimental set-up schematically. A temperature controlled external cavity laser diode (LD) actively stabilized on an atomic resonance of rubidium at 780 nm serves as light source.[18] The main laser beam is steered through two Faraday isolators (FI) to suppress light being back reflected into the laser diode and thus providing stable laser operation. Together with the entrance polarizer of the second FI the first $\lambda/2$ retarder ($\lambda/2$) allows for adjusting the beam power. Detector D2 (photodiode, BPW33) samples half of the beam to monitor the light power before it is coupled (FC) into a glass fibre and sent to the cavity (C). The light reflected from and transmitted through the cavity is collected on detectors D3 and D4 (Thorlabs, PDA 55), respectively. The electrical signal from D4 is used to stabilize the cavity, whereas the signal from D3 is sent to either a spectrum analyzer (Rohde&Schwarz, FSP 3) or a network analyzer (Rohde&Schwarz, ZVB 4).

The stabilization of the laser consists in locking the laser frequency on a reference atomic transition of Rubidium. In detail the implementation of our setting goes as follows. The temperature and current of the laser diode are controlled by a Toptica DCC 110 controller. The collimated laser beam exiting the diode (LD) is directly sent on a diffraction grating (G) which, along with the back-facet of the diode, forms an additional external cavity that narrows the line width of the laser to typically some 100 kHz.[18] Via the position of the grating the wavelength of the laser can be selected within a range spanning a few nanometers. A part of the laser beam is branched off at a glass plate (P), split in two counter propagating beams that overlap in a Rb gas cell (RB) to perform a Doppler-free absorption spectroscopy of Rb gas at detector D1 (details of such spectroscopy are not explained here but can be found in text books, e.g. reference 19). In order to lock the laser frequency, the current injected into the laser diode is modulated at 100 MHz by the reference signal of the lock-in amplifier (LI1) with the effect of modulating the laser wavelength. The signal D1 from the Rb spectroscopy is demodulated by LI1 and the resulting DC signal shows a zero point in the spectrum at the frequency of the F = 3 line of the $Rb^{85}$ isotope at 780 nm. The consecutive servo electronics (PI1), a proportional integral feedback loop, settles the laser frequency on this position by acting back via the piezo element (PZT) on the grating (G) to compensate for slow thermal drifts and on the laser current to counteract high frequency disturbances.

The fibre-based cavity itself is stabilized on the laser wavelength. To that purpose the lock-in amplifier LI2 (EG&G, 7265) modulates the cavity length by applying a 100 kHz AC voltage on the first shear piezo (PZT1) acting on the front mirror of the cavity (C). The optical signal transmitted through the cavity is sent to a detector (D4) and then demodulated by LI2. Hence the resulting DC signal has a zero point at the resonance of the cavity. A servo loop (PI2) locks on this zero point by controlling the voltage amplified by a high voltage amplifier (HV) and applied on the second shear piezo (PZT2) holding the rear mirror. The cavity is consequently kept on resonance.

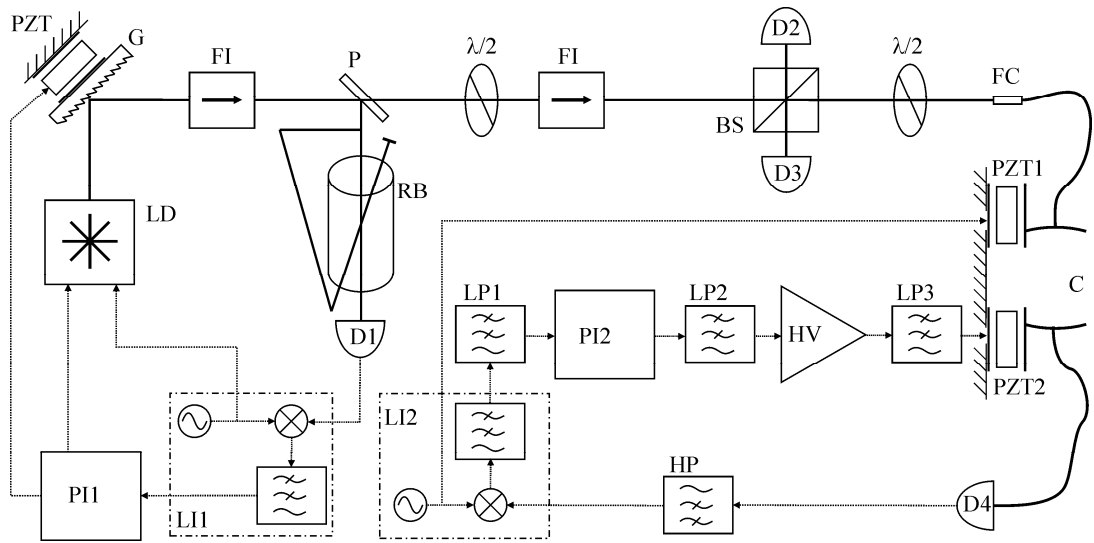

Figure 3. Schematics of the set-up as explained in the text. Thick black lines denote light path, dashed arrowed lines denote electrical signal ways. Further elements include laser diode (LD), diffraction grating (G), Faraday isolator (FI), glass plate (P), rubidium gas cell (RB), $\lambda/2$ retarder ($\lambda/2$), beam splitter (BS), fibre coupler (FC), cavity (C), detector (D), high-pass filter (HP), low-pass filter (LP), lock-in amplifier (LI) illustrated by a contour containing reference source, mixer and low-pass filter, servo electronics (PI), ±410V high voltage amplifier (HV), piezo element (PZT). Further optical beam shaping and steering elements are not drawn here.

We carefully integrated filters into the latter feedback loop line in order to suppress excess noise in the optical signal originating from servo electronics. A 48 kHz high-pass filter (HP) prevents LI2 from overload due to low frequency signals and enables operation with larger dynamic range. The noise generated at the output of LI2 is efficiently suppressed by a 4th order Bessel low-pass filter (LP1) with 2.5 kHz cut-off frequency. As shown in figure 4a) the noise spectrum of the electric signal has a lower level and less spikes after the Bessel filter (blue) than before (red). It is almost at the spectrum analyzer's internal noise level (black). Above 1 MHz the background noise is slightly elevated by the Bessel filter. Electrical noise at the output of the servo electronics (PI2) (figure 4b, red) is suppressed by a 10 kHz low-pass filter (blue). Since the high voltage signal from HV driving PZT2 is to noisy to operate the cavity properly a 16 Hz low-pass filter (LP3) is mounted at its output. Additional reduction of the cavity optical signal excess noise is achieved by using two different detectors for implementation of the cavity stabilization (D4) and for the detection of the optical signal itself (D3). Figure 4c) shows the noise spectrum of the reflected signal with 1.5 mW of optical power impinging on the cavity. The red curve corresponds first to a cavity stabilization performed using the reflected signal from D3 (red). For the blue curve on the contrary, the D4 transmission detector is used for stabilization and the Bessel-filter is integrated in the corresponding servo loop. Main peaks in the spectrum are quenched in this latter configuration and the whole noise background is reduced permitting increased sensitivity in subsequent measurements.

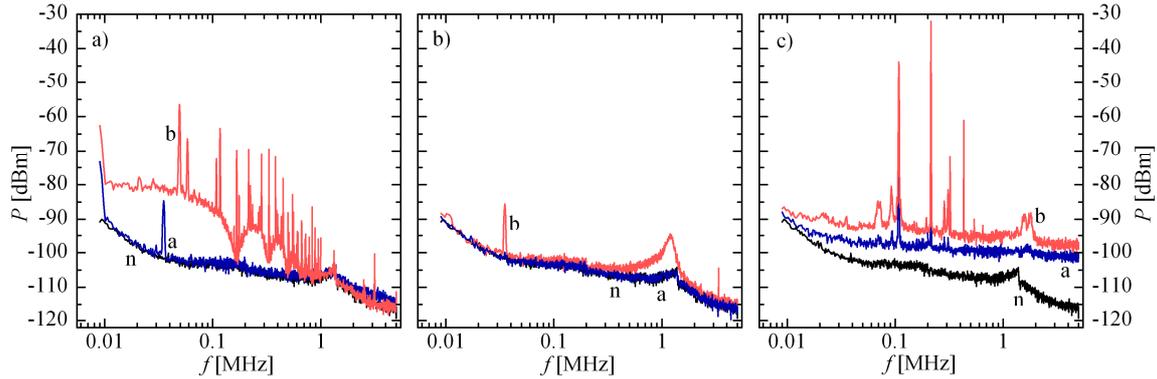

Figure 4. (color online) Noise spectra of electric and optical signals taken with 300 Hz bandwidth. Black plots labeled with (n) show the noise level of the spectrum analyzer, while red plots labeled with (b) show the noise spectra before the respective filter and blue plots labeled with (a) after the respective filter. a) Electrical noise reduction due to Bessel filter (LP1) and b) electrical noise reduction due to 10 kHz low-pass filter (LP2) in the cavity stabilization feedback loop (see also figure 3). c) Reduction of noise in the reflected optical signal due to electric filtering in the cavity feedback loop (LP1) and due to separation of cavity stabilization and cavity reflection detectors.

## 3. NANOMECHANICAL RESONATORS

In our nano-optomechanics experiment we investigate carbon-based nanomechanical resonators which have been grown by electron beam deposition (EBD) of amorphous carbon.[20] In contrast to conventional, commercially available EBD-grown nanotips for scanning probe microscopy applications, each of our nanorods extrudes from the hosting AFM-cantilever in prolongation of the lever (depicted in figure 5a) and not from the apex of the cantilever pyramid.[21] The dimensions of the cone-shaped nanorods vary between 3 and 4 µm in length and from 90 to 134 nm in diameter (for some examples see table 1 and figure 5b to e).

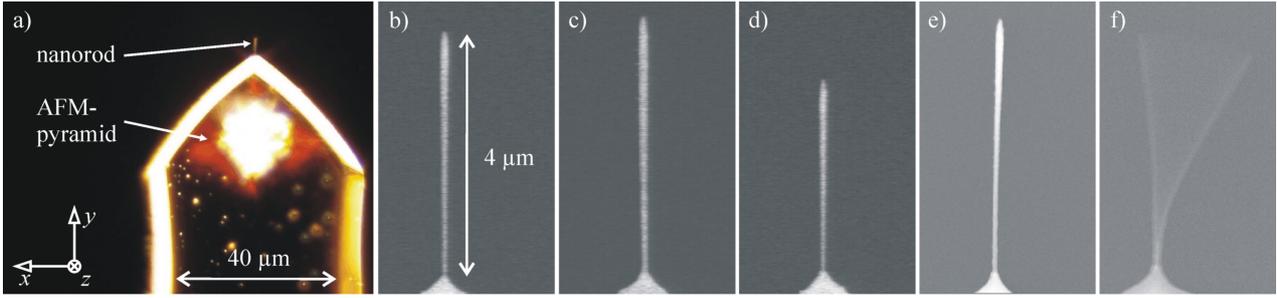

Figure 5. Optical a) and SEM images b)-f) of exemplary EBD rods. a) Optical microscope image of the free end of an AFM-cantilever holding an EBD-rod. The AFM-pyramid points along negative $z$-axis. b), c) and d) Sequence of SEM images of nanorod 21, 22 and 23 illustrating the variation in geometry occurring during fabrication (see also table 1). e) A nanorod at rest and f) the same rod mechanically excited at its resonance frequency.

To characterize the mechanical properties of these amorphous carbon rods resonant excitation experiments are performed. To this end the AFM-cantilever holding the nanorod is mounted on a thin piezo transducer which is electrically connected to a signal generator (Rohde&Schwarz, SML 02) and used to drive the mechanical motion of the nanorod. Scanning electron microscopy (SEM) imaging while sweeping the drive frequency reveals the envelope of the mechanical mode and allows to measure the rod's mechanical resonance frequencies (figure 5e and. f). However, while allowing for a detailed characterization of the rods' motion, a drawback of this technique is the deposition of further amorphous carbon on the rod during observation in the electron beam.

Table 1. Dimensions (length $l$ and diameter $2r$) of nanorods from three different batches. Resonance frequencies of their vibrational modes along $z$-axis measured in the optical microscope, together with associated quality factors.

| nanorod | $l$ [µm] | $2r$ [nm] | measured | | | | | |
|---|---|---|---|---|---|---|---|---|
| | | | $f_0$ [kHz] | $Q_0$ | $f_1$ [kHz] | $Q_1$ | $f_2$ [kHz] | $Q_2$ |
| 02 | 3.869 | 134 | 473.6 | 250 | 784 | 400 | 1172.1 | 500 |
| 11 | 3.011 | 111 | 1706.3 | 630 | 4375 | 1200 | - | - |
| 12 | 3.618 | 134 | 1072.1 | 390 | 1497.2 | 340 | - | - |
| 13 | 4.210 | 111 | 1162.5 | 360 | - | - | - | - |
| 14 | 3.896 | 134 | 1138.0 | 569 | - | - | - | - |
| 15 | 3.757 | 111 | 1300.0 | 500 | - | - | - | - |
| 21 | 3.810 | 134 | 1058.4 | 415 | 1479.0 | 565 | - | - |
| 22 | 4.036 | 134 | 980.89 | 633 | 1370.9 | 476 | - | - |
| 23 | 3.009 | 134 | 1690.9 | 421 | 2161.0 | - | 4340.8 | - |

Even though the resonators' diameter is only of the order of 100 nm, the resolution of an optical microscope is sufficient to recognize the idle nanorod with a 500-fold magnification as it is demonstrated in figures 6, 7 and 8. To avoid the above described resonator contamination in the SEM, excitation experiments were carried out under the optical microscope in ambient atmosphere. Under external actuation, broadening of the rod can still be observed in these conditions. In this case the vibrational amplitude is measured with an imprecision of 0.2 µm which originates from the limited resolution of the optical microscope.

We characterize in-plane (along $x$-axis, figure 8) and out-of-plane resonances (along $z$-axis, figure 6 and 7) of the nanorod. To observe out-of-plane resonances it is necessary to focus on the tip of the cantilever from the side. To that purpose the lever is tilted due to geometrical constraints under the microscope. The AFM-cantilever therefore points 45° out of the image plane. In this arrangement the nanorod's tip and clamping point are no longer simultaneously in focus. Note that the sketch of the AFM-lever in side view (figure 6g) is oriented the same way as on the other images (figure 6a-f, 7a-c, 8d and e) such that the pyramid of the AFM-cantilever points along the negative $z$-axis.

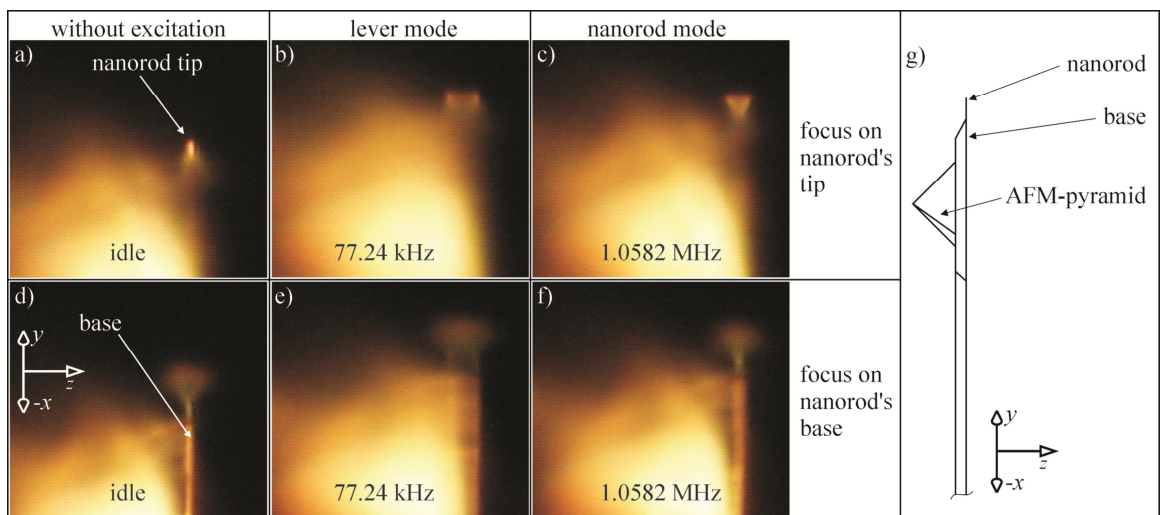

Figure 6. Side-views of actuated nanorod 21 in the optical microscope. Excitation frequencies are indicated on each picture. The $x$- and $y$-axis of the coordinate systems are tilted µ 45° with respect to the image plane whereas the lever axis is aligned with the $y$-axis. a), b) and c) focus on the free end whilst d), e) and f) focus on the clamping point of the nanorod. The excitation is switched-off in a) and d). For a mechanical resonance of the AFM-cantilever the tip b) as well as the clamping point e) of the nanorod oscillate. For resonances of the nanorod itself only the free end of the rod vibrates while the clamping point remains at rest. For better orientation in blurry images a sketch of the AFM-cantilever in side-view is given in g).

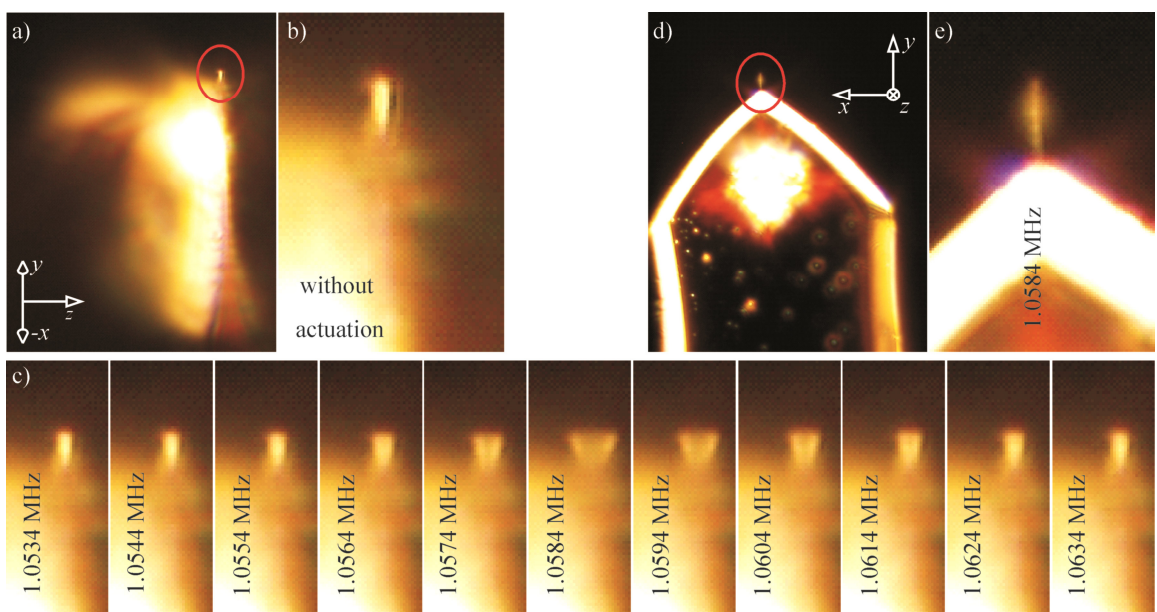

Figure 7. (color online) Optical microscope images of nanorod 21 vibrating resonantly at 1.058 MHz along the $z$-axis. a), b) and c) Side-view of the tilted AFM-cantilever apex hosting the nanorod. The red circle in a) encloses the position of the nanorod at rest. b) Close-up allows for observing vibrational amplitudes of the nanorod. The system is oriented as in figure 6a). c) Series of pictures taken during the frequency scan across a resonance illustrating the change in amplitude. The excitation frequency is indicated on each picture. d) and e) Bottom view on the AFM-cantilever and nanorod during excitation on resonance frequency at 1.0584 MHz. In the close-up the nanorod appears slightly blurred around its center due to its oscillation perpendicular to the image plane (compare with figure 8b at rest).

Despite its limited resolution the optical microscope allows to distinguish between vibrational modes of the AFM-cantilever and vibrational modes of the nanorod itself (see figure 6). If the whole cantilever resonates, the nanorod (figure 6b, focus on tip) and its clamping point (figure 6e, focus on base) are both oscillating, whereas if the drive only excites the nanorod at its own resonance, the clamping point is visually at rest (figure 6f) and the free end of the rod appears diffuse (figure 6c). For reference the nanorod and its clamping point at rest are shown in figures 6a and figure 6d, when the drive is switched off.

Scanning the drive frequency across a resonance while measuring the vibration amplitude allows to determine the quality factor $Q$ of the resonance. Due to the limited resolution of the microscope this technique implies an error of up to 30 %, but without entailing any contamination of the sample. Figure 7c) shows a series of pictures that illustrate a frequency scan across the resonance at 1.058 MHz where nanorod 21 oscillates along the $z$-axis. For comparison figure 7a) shows this rod in side-view at rest and b) the close-up view used during the scan. Observed with a perpendicular bottom view, this vibration appears as a slight blurring around the nanorod's centre as depicted in figure 7d) and e). Additionally a hysteresis is observed when scanning the frequency up and down across the resonance revealing anharmonic, non-linear behaviour.[22] Due to large motional amplitudes required to optically resolve the resonances with the described technique, some of our characterization measurements had to be performed in this nonlinear regime. At 1.677 MHz nanorod 21 resonates along the $x$-axis. Figure 8a) and b) depict the nanorod in bottom view at rest in comparison to figure 8c) which shows an image series presenting the scan across that resonance. In side-view (figure 8d and e) the nanorod seems slightly blurred when being driven at 1.6773 MHz compared to figure 7a) and b) where the drive is switched off. The amplitude information obtained from scans shown in figure 7c) and 8c) is plotted in figure 9a) and c), respectively. Figure 9b) shows a plot of an additional vibrational resonance along the $z$-axis of nanorod 21 at 1.479 MHz. Lorentzian fits in plots of figure 9 give an estimate of the quality factor of each resonance. When the resonance is asymmetric as shown in figure 9a) and 9b), the nanorod response is non-linear and a simple lorentzian fit is normally not sufficient to infer the mechanical $Q$ factor. However each $Q$ value obtained this way in our experiments is confirmed independently by a subsequent in-cavity measurement within an imprecision of ±30 %. This error is comparable to the error induced by the limited resolution of the optical microscope itself. We therefore conclude that the error in the $Q$ estimation with the optical microscope lies in this range. Table 2 provides a complete list of vibrational resonances of nanorod 21 with their $Q$ factors.

The purpose of the piezo-excitation characterization technique is to find the resonance frequencies of a given nanorod before subsequent in-cavity measurements. Despite rather large errors in amplitude the resonance frequency can be determined precisely.[23] Therefore, the characterization method in the optical microscope is preferred to the SEM since it does not contaminate the nanorod, and does not modify the resonance frequency during observation. However, since the optical microscope characterization is typically performed in air, we consistently observe lowered mechanical quality factors $Q$ due to gas damping effects.[24]

A first intuitive way to model the mechanical modes of a nanorod is to use elasticity theory applied to a singly clamped cylinder made out of an homogeneous elastic material with Young modulus $E$ and density $\rho$. In this case we have[25, 26]

$$f_n = \frac{1}{4\pi}\sqrt{\frac{E}{\rho}}\frac{r}{l^2}(\beta_n l)^2 \qquad (1)$$

for the resonance frequency $f_n$ of the $n^{\text{th}}$ mode of the cylinder with length $l$, radius $r$ and where the product $(\beta_n l)$ is a numerical value without unit. For the first three modes the respective values are $\beta_0 l = 1.8751$, $\beta_1 l = 4.6941$ and $\beta_2 l = 7.8458$. However, this model is unable to fit the observed nanorod resonances, be it by postulating the elastic constants or by trying to fit the resonances with adjustable elastic parameters. In a second attempt to simulate the nanorod's mechanical modes we used a finite element method software (Comsol Multiphysics) and the nanorod geometry was taken to be a cone instead of a simple cylinder. However this approach was not more successful. We conclude that the carbon-based nanorod material is not correctly approximated by a homogeneous elastic material and we anticipate inhomogeneous mass distribution and defects resulting from the fabrication process to be responsible for this discrepancy.

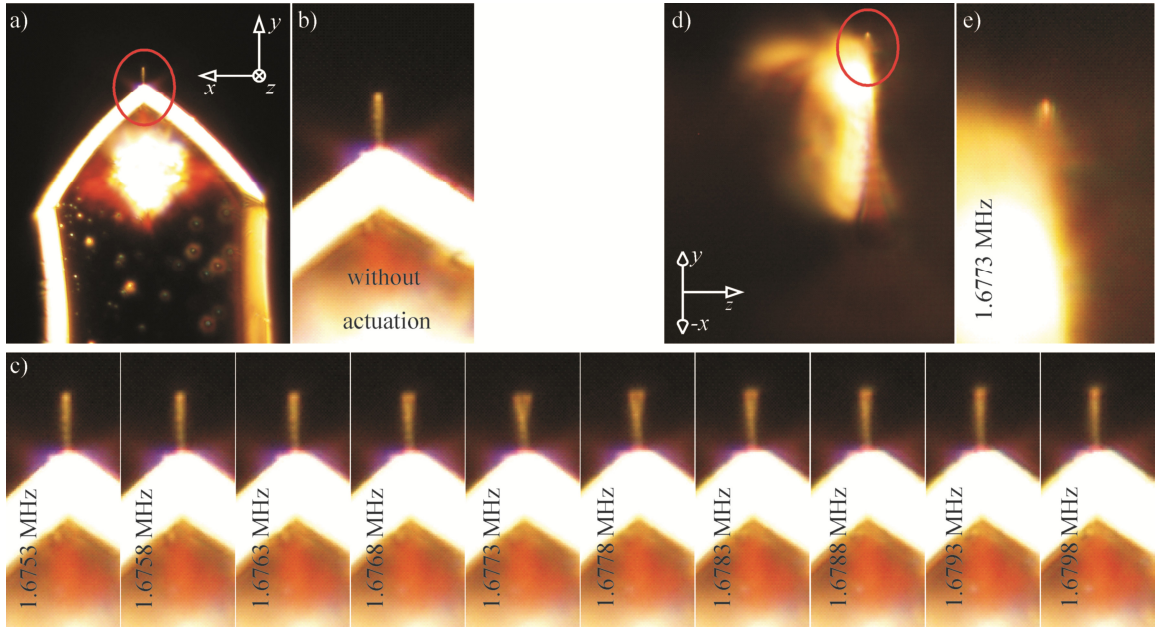

Figure 8. (color online) Optical microscope images of nanorod 21 vibrating resonantly at 1.677 MHz along the *x*-axis. a), b) and c) Bottom view of the AFM-cantilever apex hosting the nanorod. The red circle in a) encloses the position of the nanorod at rest. b) Close-up of the nanorod at rest. c) Series of pictures taken during the frequency scan across a resonance illustrating the change in amplitude. The excitation frequency is indicated on each picture. d) and e) Side-view of the AFM-cantilever and nanorod during excitation on resonance frequency at 1.6773 MHz. In the close-up e) the nanorod appears slightly blurred due to its oscillation perpendicular to the image plane (compare to figure 7b at rest). The system is oriented as in figure 6a).

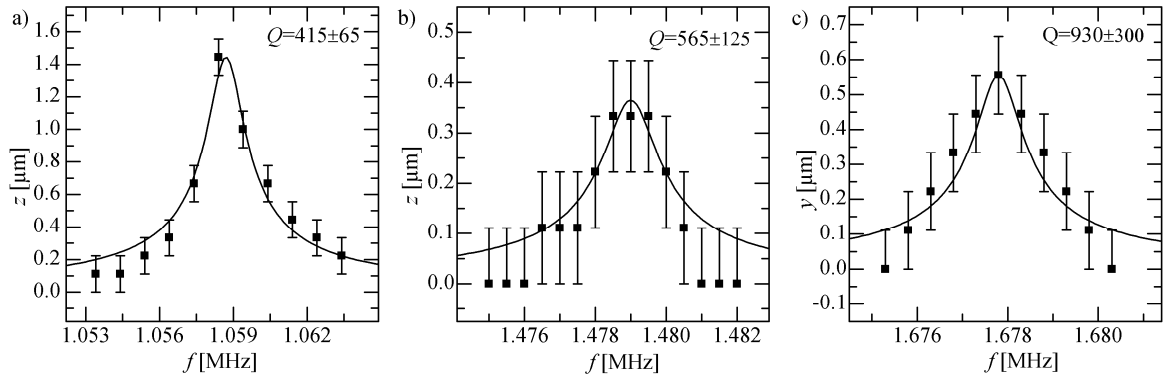

Figure 9. Resonances of nanorod 21 measured in the optical microscope with fitted Lorentzians. The error bars are set by the resolution limit of the microscope. a) Resonance along the *z*-axis at 1.0584 MHz obtained from the series shown in figure 7c). Nonlinear response of the rod is apparent from the asymetry of the resonance curve. b) Consecutive resonance along *z*-axis observed on nanorod 21. Data points are slightly asymmetrically distributed around the resonance at 1.479 MHz indicating nonlinear response. c) Resonance along *x*-axis at 1.673 MHz where data points obtained from series in figure 8c) are distributed symmetrically. Lorentzian fit gives $Q = 930$ with 30% error.

# 4. NANOMECHANICAL RESONATOR IN THE CAVITY

After measuring the nanorods' resonance frequencies, rod 21 is selected and introduced into the cavity. This is accomplished by mounting the hosting AFM-cantilever with the actuating piezo on the xyz-positioning unit for precise positioning within a 5 mm range. Coarse placement of the AFM-cantilever in the 37 µm wide cavity gap is carried out under visual observation with a microscope as shown in figure 2c). In a second, fine positioning step the cavity transmission and reflection are monitored carefully to ensure optimum positioning of the nanorod. Since the nanorod as well as the cantilever increase the loss of the cavity as soon as they plunge into the optical mode it is important to have the nanorod inserted into the cavity mode, while avoiding additional loss contributions from the cantilever. If the nanorod alone perturbs the optical cavity mode, the cavity transmission and reflection change periodically as a function of the position of the nanorod along the cavity axis.[3] A more detailed description on the positioning of the nanorod inside the cavity can be found in previous work.[4] The absorptive optomechanical coupling is maximized when the nanorod is positioned on the steepest gradient of the reflection $dR/dz_0$.[3] At this position, the optical detection of the nanorod's mechanical vibration is efficiently performed analysing the cavity transmission or reflection noise.

After positioning the nanorod as described, we use a network analyzer (Rohde&Schwarz, ZVB 4) to excite the nanorod's mechanical motion with the piezo actuator and analyse the optical signal reflected from the cavity with a resolution of 100 Hz. Figure 10 depicts the resulting vibrational spectrum of the nanorod measured in nitrogen at atmospheric pressure (black graph) and in vacuum at $10^{-5}$ mbar (red graph). Most of the peaks shown in this plot can be assigned to resonances observed previously during piezo-excitation experiments under the optical microscope (table 1 and 2). The inset in figure 10 shows a close-up at resonance $k$ with Lorentzians (blue) fitted to the respective resonances in nitrogen and vacuum. At room pressure both frequency and $Q$ factor are clearly shifted to lower values compared to vacuum. Table 2 lists the resonance frequencies and $Q$ values of all the peaks shown in figure 10 and confirms this behaviour: in nitrogen at ambient pressure frequencies go down up to one percent while quality factors decrease to approximately 10% of the value measured in vacuum. $Q$ values obtained in nitrogen confirm those observed in air under the optical microscope within the measurement imprecision. The observed behaviour can be attributed to virtual mass added by the gas which decreases the resonance frequency and to an increased influence of gas damping on dissipation of the nanomechanical resonator which lowers $Q$.[24]

Resonance $g$ and $h$ at 0.43 and 0.71 MHz have been identified previously as higher order modes of the cantilever, while $i$, $k$ and $l$ are modes from the nanorod itself at 1.06, 1.48 and 1.68 MHz. Interestingly resonance $l$ at 1.68 MHz deflects the nanorod in the $x$ direction but it is still detected 18 dB above the noise floor. However the cavity detection scheme is expected to be less sensitive to nanomechanical modes in the $x$ direction. Two effects may contribute to the detection of resonance $l$. First, the lateral energy density gradient of the cavity mode in the $xy$-plane allows to measure deflections in $x$ and $y$ directions. Second, the rod may display residual deflection along the cavity axis $z$ due to slight misalignment. Although resonance $m$ at 1.97 MHz reveals the highest $Q$ in vacuum we are not able to observe it using the piezo-exitation characterization method in the optical microscope.
Figure 10 proves the efficiency of our cavity-assisted detection method. Piezo-excited nanomechanical motion in vacuum is measured 50 dB over the noise floor in this example. Most important, this cavity optical interferometer can in principle allow optical vibrational spectroscopy of any nanomechanical system as will be shown in future work.

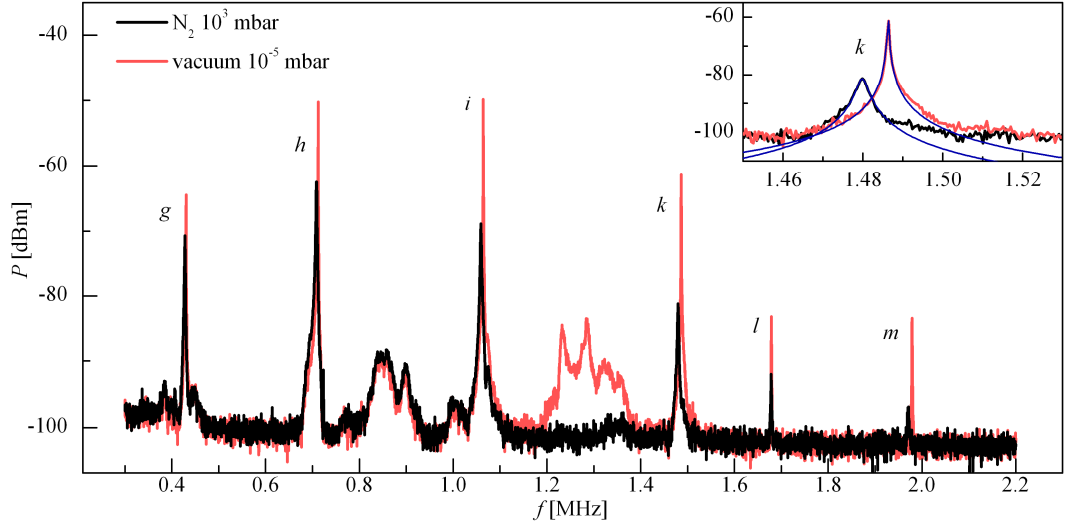

Figure 10. (color online) Optical cavity reflectivity noise spectrum showing the vibrational spectrum of the actuated cantilever-nanorod system when the nanorod is positioned in the cavity mode. Measurements are carried out in nitrogen at atmospheric pressure (black line) and in vacuum at $10^{-5}$ mbar (red line) with a bandwidth of 100 Hz. The inset gives a close-up on resonance $k$ showing different response in $N_2$ and vacuum. The blue lines are Lorentzian fits to the respective resonances. Table 2 lists the parameters of each resonance peak shown here.

Table 2. Resonances of sample 21 measured for the AFM-cantilever and the nanorod in the optical microscope in air and in the cavity in nitrogen at ambient pressure and in vacuum at $10^{-5}$ mbar. Resonances are assigned to the lever (L) or to the nanorod oscillating along the $x$- and $z$-axis (Nx and Nz) or to none of them (U) as indicated in the second line.

| index $r$ | $a$ | $b$ | $c$ | $d$ | $e$ | $f$ | $g$ | $h$ | $i$ | $k$ | $l$ | $m$ |
|---|---|---|---|---|---|---|---|---|---|---|---|---|
| type | L | L | L | L | L | L | L | L | Nz | Nz | Nx | U |
| resonances measured in the optical microscope in air | | | | | | | | | | | | |
| $f_{ar}$ [kHz] | 12.17 | 77.24 | 82.63 | 217.72 | 277 | 289 | 427.62 | 707.73 | 1058.4 | 1479.0 | 1678 | - |
| $Q_{ar}$ | - | 97 | - | 330 | - | - | 202 | 313 | 415 | 565 | 930 | - |
| resonances measured in the cavity in nitrogen at atmospheric pressure | | | | | | | | | | | | |
| $f_{nr}$ [kHz] | - | - | - | - | - | - | 427.73 | 708.52 | 1059.2 | 1479.8 | 1678.4 | 1970.9 |
| $Q_{nr}$ | - | - | - | - | - | - | 188 | 515 | 565 | 597 | 785 | 415 |
| resonances measured in the cavity in vacuum at $10^{-5}$ mbar | | | | | | | | | | | | |
| $f_{vr}$ [kHz] | - | - | - | - | - | - | 430.20 | 712.09 | 1064.2 | 1486.3 | 1678.9 | 1978.9 |
| $Q_{vr}$ | - | - | - | - | - | - | 1772 | 4354 | 4266 | 4329 | 4065 | 6314 |

## 5. SUMMARY

We present the technical realization of a set-up employing a high finesse fibre based micro cavity to optically map out driven response of mechanical nanoresonators. Both the optical cavity and the nanomechanical resonator are carefully characterized. We apply EBD grown carbon rods as model resonators in the experiment. The rods' resonances found by mechanical characterisation are recovered in the optomechanical set-up with a much higher precision. In a previous work, the Brownian motion of a nanoresonator was measured using the cavity.[4] In principle this set-up combining a miniature Fabry-Pérot cavity with a nanomechanical resonator is applicable to studying light-induced backaction.[3] Even more as shown here, it is an efficient tool to transduce nanomechanical motion into an optical signal, its versatility making it applicable to any kind of nanomechanical resonator. Such cavity nano-optomechanical system should allow fine optical spectroscopy of various interesting vibrating nano-systems for nanomechanics applications.

## ACKNOWLEDGMENT

We gratefully acknowledge financial support of the Alexander von Humboldt Foundation, the German-Israeli Foundation (G.I.F.), the German Excellence Initiative via the Nanosystems Initiative Munich (NIM), the Center for NanoScience (CENS) and the DAAD/Egide Procope (German/French) exchange program.